\documentclass[aps,prl,twocolumn,showpacs]{revtex4}
\usepackage{times}
\usepackage{graphicx}
\usepackage{graphics}
\usepackage{dcolumn}
\usepackage{amsmath}
\def\lambdabar{\protect\@lambdabar}
\def\@lambdabar{%
\relax \bgroup
\def\@tempa{\hbox{\raise.73\ht0
\hbox to0pt{\kern.25\wd0\vrule width.5\wd0
height.1pt depth.1pt\hss}\box0}}%
\mathchoice{\setbox0\hbox{$\displaystyle\lambda$}\@tempa}%
{\setbox0\hbox{$\textstyle\lambda$}\@tempa}%
{\setbox0\hbox{$\scriptstyle\lambda$}\@tempa}%
{\setbox0\hbox{$\scriptscriptstyle\lambda$}\@tempa}%
\egroup }
\begin{document}

\title{Parametric Amplification of Atoms}

\author{Gretchen K. Campbell, Jongchul Mun, Micah Boyd, Erik W. Streed, Wolfgang Ketterle and David E. Pritchard}
\homepage[URL: ]{http://cua.mit.edu/ketterle_group/}
\affiliation{MIT-Harvard Center for Ultracold Atoms, Research
Laboratory of Electronics and Department of Physics, Massachusetts
Institute of Technology, Cambridge, MA 02139, USA}
\date{\today}

\begin{abstract}
We have observed parametric generation and amplification of
ultracold atom pairs. A $^{87}$Rb Bose-Einstein condensate was
loaded into a one-dimensional optical lattice with quasimomentum
$k_{0}$ and spontaneously scattered into two final states with
quasimomenta $k_{1}$ and $k_{2}$ . Furthermore, when a seed of
atoms was first created with quasimomentum $k_{1}$ we observed
parametric amplification of scattered atoms pairs in states
$k_{1}$ and $k_{2}$ when the phase-matching condition was
fulfilled. This process is analogous to optical parametric
generation (OPG) and amplification (OPA) of photons and could be
used to efficiently create entangled pairs of atoms. Furthermore,
these results explain the dynamic instability of condensates in
moving lattices observed in recent experiments.

\end{abstract}

 \maketitle

Nonlinear atom optics is a novel research area born with the advent
of Bose-Einstein condensates of alkali atoms. Unlike photons,
ultracold atoms have a very strong nonlinearity directly provided by
s-wave collisions, and therefore they do not need a nonlinear medium
to provide effective interaction. A number of nonlinear processes
first observed with photons have been demonstrated with matter waves
such as four-wave mixing \cite{Deng99,Vogels02}, solitons
\cite{Denschlag00,Burger99,Anderson01,Strecker02}, second-harmonic
generation \cite{Wynar00, Abo-Shaeer05, Lett93, Miller93}, and
sum-frequency generation \cite{Abo-Shaeer05}. Nonlinear atom optics,
and in particular four-wave mixing, has previously been suggested as
an ideal way to create entangled pairs of atoms
\cite{Duan00,Pu00,Vogels02}. However in previous four wave mixing
experiments \cite{Deng99,Vogels02} using condensates in free space,
the quadratic dispersion relation for free particles only allowed
for the phase-matching condition to be fulfilled when the magnitudes
of all four momenta were equal (in the center-of-mass frame). This
is the only way in which two particles can scatter off each other
and conserve energy and momentum. In particular, in free space, if a
condensate is moving with momentum $k_{0}$, atoms within the
condensate cannot elastically scatter into different momentum
states, and therefore the analog to optical parametric generation of
photons is not possible.

The situation is very different, when an optical lattice is added.
The lattice delivers energy in the form of the AC Stark effect and
momentum in units of $2\hbar k_L$ to the atoms, where $k_L$ is the
wavevector of the optical lattice. The motion of atoms in this
periodic potential is described by a band structure, which
deviates from the quadratic free particle dispersion curve. In a
lattice, as recently  suggested \cite{Hilligsoe05}, it becomes
possible for two atoms in the condensate to collide and scatter
into a higher and lower quasimomentum state and conserve energy.
As we discuss below, this can lead to dynamic instabilities of the
condensate, but also enables non-degenerate four-wave mixing and
the atom-optics analog of optical parametric generation.

Phase matching is essential for high efficiency in nonlinear
processes in quantum optics including optical parametric generation
of photons \cite{Rosencher02}. A modification of the dispersion
curve has been used to demonstrated optical parametric amplification
in semiconductor microcavities \cite{Savvidis00}.  In atom optics,
dispersion management was used to modify the effective mass of atoms
\cite{Eiermann03}, and to create bright gap solitons
\cite{Eiermann04}. Here we demonstrate that by modifying the
dispersion curve using an optical lattice scattering processes which
cannot occur in free space become possible, and realize the
matter-wave analogue of an Optical Parametric Generator (OPG) and an
Optical Parametric Amplifier (OPA).

\begin{figure}
\centering{
\includegraphics[hiresbb=true,width=8cm]{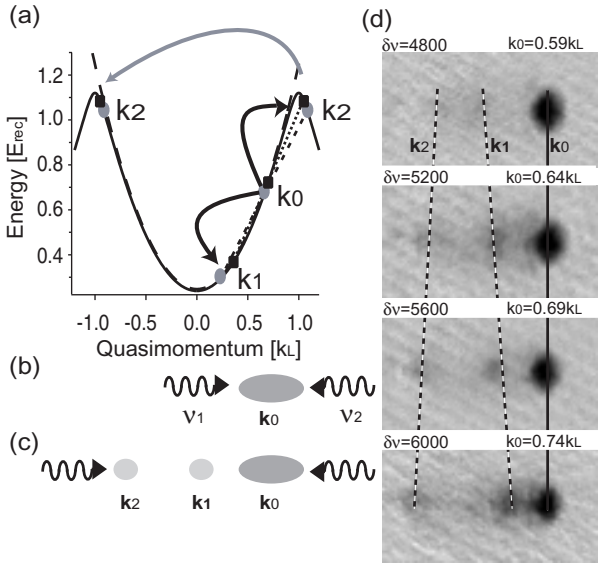}
\caption{\label{fig:figone} Dispersion curve for the optical lattice
and experimental setup. (a) Band structure for a lattice depth of
V=0.5$E_{rec}$. The dashed line shows the free particle dispersion
curve. The dispersion relation of the lattice allows two atoms with
momentum $k_0$ to elastically scatter into the final momentum states
$k_1$ and $k_2$.  Energy and quasimomentum are conserved when $k_0$
is the average of $k_1$ and $k_2$ and the three points on the
dispersion curve lie on a straight line. If $k_0$ is varied, the
allowed values for $k_1$ and $k_2$ change. For values of $k_0$ below
$\approx$ 0.55$k_L$, where $k_L$ is the wavevector of the optical
lattice, atoms cannot scatter elastically into different momentum
states. The circles (squares) show allowed states $k_0$,$k_1$,$k_2$
for $k_0=$0.66$k_L$ ($0.70k_L$). As $k_{0}$ is increased, the final
momentum states move closer together. Since the scattering occurs
within the lowest band of the lattice, the final momentum is
$k_{2}=\left(2k_{0}-k_{1}\right)Mod(2k_{L})$ (b) A $^{87}$ Rb
Bose-Einstein condensate is illuminated by two counter-propagating
laser beams with detuning $\delta\nu$, which create a moving optical
lattice. The condensate is initially held at rest in a magnetic
trap. In the rest frame of the lattice, the condensate has
quasimomentum $k_{0}=\frac{m\lambda}{2\hbar}\delta\nu$. (c) As the
quasimomentum $k_0$ of the condensate was varied, we observed
elastic scattering into states $k_1$ and $k_2$. (d) Absorption
images for different lattice detunings, $\delta\nu$ showing
parametric generation. After ramping up the lattice, the atoms were
held for 10 ms at a constant lattice depth. They were then released
from the trap and imaged after 43 ms of ballistic expansion. The
field of view is 0.5 mm$ \times $0.3 mm}}
\end{figure}

To demonstrate the matter-wave analogue of an OPG, a $^{87}$Rb
Bose-Einstein condensate with quasimomentum $k_{0}$ was loaded into
a one-dimensional optical lattice. To load the atoms at a given
quasimomentum relative to the Brillouin zone, a one-dimensional
moving optical lattice was adiabatically applied to a magnetically
trapped condensate initially at rest in the lab frame. The lattice
was created using two counter-propagating laser beams with frequency
difference $\delta\nu$, giving the lattice a velocity of
$v=\frac{\lambda}{2}\delta\nu$, where $\lambda$ is the wavelength of
the optical lattice. In the rest frame of the lattice, the
condensate has quasimomentum
$k_{0}=\frac{m\lambda}{2\hbar}\delta\nu$, where m is the atomic
mass. By changing the detuning between the lattice beams,
$\delta\nu$, the quasimomentum $k_{0}$ of the condensate could be
varied. As shown in Fig.~\ref{fig:figone}d, as the value of $k_{0}$
was varied we observed elastic scattering of atom pairs into final
states $k_{1}$ and $k_{2}$.  The range of possible final states
varied with $k_{0}$ due to the phase matching condition. For values
of $k_{0}$ less than $\approx.55k_{L}$  the dispersion relation
imposed by the Bloch structure of the optical lattice does not allow
elastic scattering to occur. For our lattice depth of $V$ =
0.5$E_{rec}$, where $E_{rec}$ = $\hbar^2 k_L^2/2m$, the values of
$k_{2}$ which satisfied energy and momentum conservation were beyond
the first Brillouin zone. Since the scattering process occurs within
the first Bloch band, the atoms in state $k_{2}$ have a
quasimomentum $k_2$ = $\left(k_{0}-k_1\right)Mod(2 k_{L})$ (see
Fig.~\ref{fig:figone}a). As the value for $\delta\nu$ (and the
resulting value of $k_{0}$) is increased, the separation between
$k_{0}$ and the allowed quasimomentum states $k_{1}$ and $k_{2}$
decreases as clearly observed in Fig.~\ref{fig:figone}d. For values
of $k_{0}$ above $\approx$ 0.75$k_L$, the final momentum states were
no longer distinguishable, and the condensate became unstable.

\begin{figure} \centering{
\includegraphics[hiresbb=true,width=8cm]{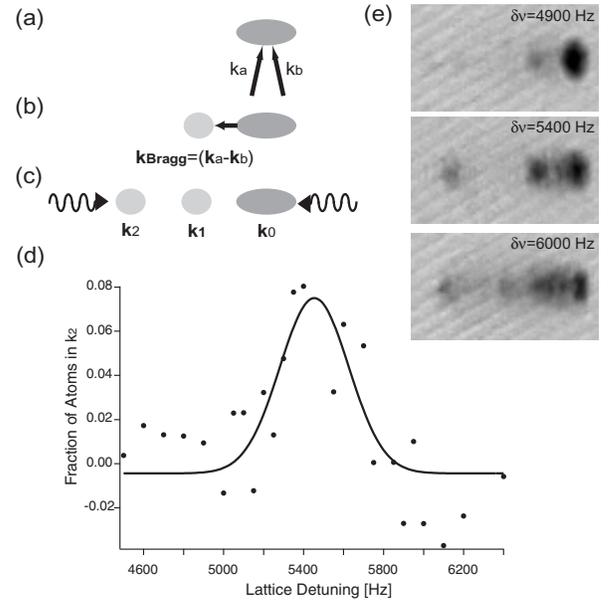}
\caption{\label{fig:figtwo}  Parametric amplification of scattered
atom pairs in a 1D optical lattice. (a) First, a 2 ms Bragg pulse
was applied to the condensate. (b) The Bragg pulse seeded atoms
along the long axis of the condensate with momentum $k_{Bragg}$ =
$(k_a-k_b)$ in the lab frame (c) The optical lattice was then
adiabatically ramped on and applied for 10 ms. When the phase
matching condition was fulfilled, parametric amplification of atoms
in the seeded state $k_1$ and its conjugate momentum state $k_2$ was
observed. (d) Resonance curve showing amplification of $k_2$, when
$k_1$ is seeded. Amplification occurred only when the phase-matching
condition was met. For a fixed $k_{Bragg}$, the resonance condition
was found by varying the detuning $\delta \nu$ of the lattice. The
data was taken for $k_{Bragg}$ = 0.43$k_{L}$. The fraction of
amplified atoms was obtained by subtracting images with and without
the seed pulse.  (e) Absorption images showing amplification of
$k_1$ and $k_2$ when the phase matching condition is met. The center
of the resonance was at $\delta\nu\approx$ 5450Hz, close to the
calculated value of $\delta\nu\approx5350Hz$.}}
\end{figure}

\begin{figure}
\centering{
\includegraphics[hiresbb=true]{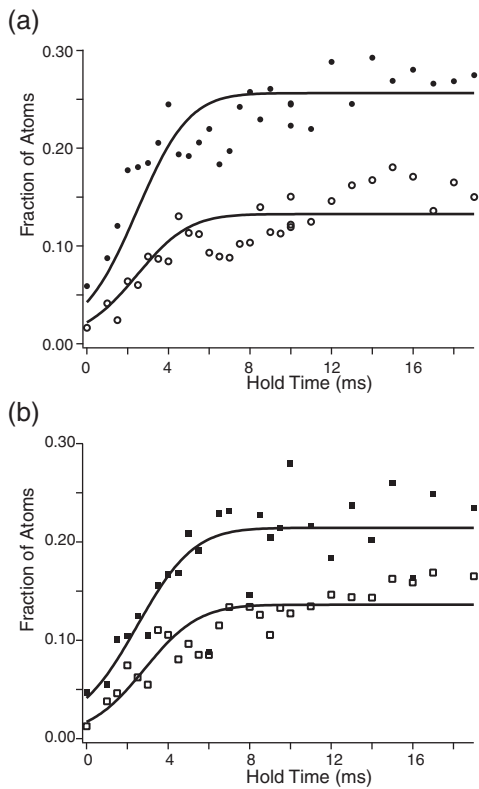}
\caption{\label{fig:figthree} Growth curve for atomic population in
quasimomentum states $k_{1}$ and $k_{2}$ when the process is seeded.
(a) Amplification of atoms with quasimomentum $k_{1}$ (solid
points), and with the conjugate momentum $k_{2}$ (open points), when
state $k_{1}$ was seeded. (b) Amplification of atoms with
quasimomentum $k_{1}$ (solid points), and with the conjugate
momentum $k_{2}$ (open points), when state $k_{2}$ was seeded. The
values for $k_{0}$,$k_{1}$,$k_{2}$ were 0.66$k_{L}$,0.23$k_{L}$,and
-0.92$k_{L}$ respectively. The solid lines shows the expected gain
using Eq.~(\ref{eq:one}) with a variable scale factor as the only
free parameter.}}
\end{figure}

To demonstrate the matter wave analogue of optical parametric
amplification, we first created a small seed of atoms with
quasimomentum $k_{1}$ before ramping on the moving lattice (see
Fig.~\ref{fig:figtwo}).
To create the seed we applied a Bragg pulse to the magnetically
trapped condensate, outcoupling a small fraction of atoms into the
momentum state $k_{Bragg}$ \cite{Stenger99}. Immediately after
applying the pulse, the optical lattice was adiabatically ramped on.
In the rest frame of the lattice, the seed has quasimomentum
$k_{1}=k_{Bragg}+k_{0}$. The phase matching condition for a given
seed momentum $k_{Bragg}$ was found by varying the frequency
difference $\delta\nu$ of the lattice, and therefore the
quasimomenta $k_{0}$ and $k_{1}$ of the atoms. As shown in
Fig.~\ref{fig:figtwo}d, when the phase matching condition was
fulfilled, we observed amplification of the seed $k_{1}$ as well as
its conjugate momenta $k_{2}$. The growth of the seed and its
conjugate momentum as a function of time are shown in Fig.
~\ref{fig:figthree}.

The experiments were performed using an elongated $^{87}$Rb
Bose-Einstein condensate created in a cloverleaf-type
Ioffe-Pritchard magnetic trap previously described in Ref.
\cite{Schneble03}. The magnetic trap had a radial (axial) trap
frequency of 35(8) Hz. The condensate, containing between
$0.5-3.0\times10^5$ atoms, was produced in the $|5^2S_{1/2},F=1,
m_F=-1\rangle$ state. The Bragg pulse was created with two laser
beams derived from the same laser, which was red-detuned from the
$5^2S_{1/2},F=1\to 5^2P_{3/2},F=1$ transition at $\lambda=780$nm by
400MHz, and was $\pi$-polarized. As shown in Fig ~\ref{fig:figtwo},
the Bragg beams were aligned such that atoms were outcoupled along
the long axis of the condensate. The intensity of the Bragg pulse
was chosen such that less than 5\% of the initial condensate was
outcoupled into the state $k_{Bragg}$, and the length of the pulse
was 2ms. The angle between the Bragg beams could be varied to change
the momentum of the outcoupled atoms. The optical lattice was
created using two counter-propagating beams derived from the same
laser with $\lambda=1064$nm, and the frequency of the two beams were
controlled by two separate acousto-optic Modulators (AOMs) driven
with a frequency difference $\delta\nu$. The lattice was also
aligned along the long axis of the condensate, and was ramped on in
1 ms using an exponential ramp.  After the condensate was held in
the lattice for a variable time $\tau$ it was then released from the
trap and imaged after 43 ms of ballistic expansion.

For all of our experiments, the depth of the optical lattice was V =
0.5 $E_{rec}$ with a band structure shown in Fig. ~\ref{fig:figone}.
When the process was not seeded, atoms were elastically scattered
into a narrow band of states $k_{1}$ and $k_{2}$, where both energy
and momentum was conserved.  However the population in neither state
was large enough for amplification to be observed. When the process
was seeded, amplification occurred when the quasimomentum was tuned
such that energy and momentum were conserved for the states $k_{0}$,
$k_{1}$, and the conjugate momentum $k_{2}$. In our experiment, the
difference $\Delta k$ = $k_{0}-k_{1}$ between the quasimomentum of
the condensate $k_0$ and seed $k_1$ was set by the angle of the
initial Bragg pulse. In the rest frame of the lattice, for a given
frequency difference $\delta\nu$ the quasimomentum of the atoms was
$k_{0}=\frac{m\lambda}{2\hbar}\delta\nu$, and that of the seed was
$k_{1}=k_{Bragg}+k_{0}$.

For a given Bragg angle, there is only one set of quasimomenta
$k_{0}$,$k_{1}$, and $k_{2}$ where the phase-matching condition is
fulfilled. To find this point, we varied the velocity of the moving
lattice for fixed hold times. Results for $k_{Bragg}$=0.43$k_{L}$
are shown in Fig.~\ref{fig:figtwo}. The phase-matched value for
$k_{2}$ is at 1.08$k_{L}$, beyond the boundary of the first
Briollouin zone. Therefore, the atoms are observed with a momentum
$k_2$ = -0.92$k_{L}$. For $k_{Bragg}$ = 0.43$k_L$,0.34$k_L$ and
0.28$k_L$, we observed resonances at $\delta\nu$=5450Hz, 5750 Hz and
6100Hz respectively. For these Bragg angles and our lattice depth,
the expected values were 5350Hz, 5700Hz and 6050Hz.

In Fig~\ref{fig:figthree}, 5\% of the initial condensate containing
$N_0$ = 1.3(3)$\times10^{5}$ atoms was outcoupled with $k_{Bragg}$ =
0.43$k_{L}$. The gain for the process is determined by the strength
of the nonlinear interaction $U$ between atoms in the condensate.
For a shallow lattice $U$ = $\frac{4\pi\hbar^2a}{m}$, where a is the
s-wave scattering length. From this we can estimate the maximum
amplification rate to be $\eta$ = $2n_0U/\hbar$ \cite{Vogels02},
with $\dot{N}_{1(2)}$ = $\eta N_{2(1)}$, where $N_{1(2)}$ is the
number of atoms in the momentum state $k_{1(2)}$. For $N_0$ =
1.3(3)$\times10^{5}$, the maximum growth rate should be $\eta$ = 540
Hz. The amplification rate will decrease as the state $k_0$ is
depleted.  However, for our small seeds, the amplification was
limited by the loss of overlap between the condensate and the
amplified pulses.  The Thomas-Fermi radius ($R_{TF}$)of the
condensate in the axial direction was 33$\mu$m, and the recoil
velocity ($v_{rec}$) for the final states $k_1$ and $k_2$ with
respect to the initial condensate was $v_{rec}$=1.8 $\mu$m/ms and
6.8$\mu$m/ms respectively. The overlap integral between the
amplified atoms and the initial condensate can be approximated as a
Gaussian with time constant $\tau_c\approx0.75R_{TF}/v_{rec}$, which
for our parameters is 3.75 ms. We compare our results to the
modified rate equation
\begin{equation}
\label{eq:one} \dot{N}_{2(1)}=\eta N_{1(2)}
e^{\frac{-t^2}{\tau_c^2}}
\end{equation}
Since atoms are scattered into states $k_1$ and $k_2$ in pairs, one
would expect that the final atom number in the two states (minus the
initial seed) are equal.  Instead, we observe a smaller number in
state $k_2$ which we ascribe to the proximity of $k_2$ to the
boundary of Brillouin zone.  This leads to instabilities, where
atoms in state $k_2$ are scattered into other momentum states or
into higher bands. If we allow a variable scale factor to correct
for this in our model, as shown in Fig.~\ref{fig:figthree}, the
final values for $N_{1,2}$ are in agreement with the experimental
data.

Amplification was also observed when atoms were seeded in state
$k_{2}$. Due to the geometry of our experimental setup, we were
unable to load atoms directly into $k_{2}$ = -0.92$k_{L}$. However
when atoms with quasi-momenta $k$ = 1.08$k_{L}$ were loaded into the
lattice, the ramp-up was no longer adiabatic due to their proximity
to the boundary of the first Brillouin zone. Because of this, atoms
from the seed were loaded into both the second Bloch band (with $k$
= 1.08$k_{L}$) and the ground state (with $k$ = -0.92$k_{L}$ =
$k_{2}$). As shown in Fig.~\ref{fig:figthree}b, the gain for this
process was almost identical to when atoms were seeded in state
$k_{1}$.

The loss of overlap could be alleviated by using a more extreme
trap geometry in which the condensate is more elongated, e.g. by
confining atoms in a tight transverse optical lattice. In this
configuration, it may be possible to observe the parametric
scattering dynamics for longer time scales, which may allow for
the observation of Rabi oscillations between $k_0$ and $k_1$,
$k_2$ as predicted in Ref.~\cite{Hilligsoe05}. For longer
coherence times, parametric amplification could also be an
efficient means of producing pairs of momentum entangled atoms for
quantum information applications \cite{Duan00,Pu00}.

For high atom numbers, and for large values of $k_0$ the condensate
became unstable, and scattered into a broad band of final momentum
states (Fig.~\ref{fig:figtwo}e). For $k_{Bragg}$ = 0.43$k_{L}$, the
energy of atoms outcoupled by the Bragg beams was $\approx 370Hz$,
whereas the chemical potential of the condensate was $\approx300Hz$.
Because of this, if the atom number was increased significantly the
momentum peaks were no longer distinguishable. When the chemical
potential of the condensate was larger than the separation between
the phase matched momentum states, the process was self-seeded, i.e
the momentum spread of the initial condensate contained atoms with
momentum $k_{0}$,$k_{1}$,and $k_{2}$, and considerable scattering
occurred. Similarly, if the atom number was kept constant, and the
value of $k_{0}$ was increased, the phase-matched momentum states
move closer together until they are no longer distinguishable.  This
occurred at values of $k_{0}$ above $\approx$ 0.75$k_{L}$, and we
observed a dynamic instability. For larger atom numbers, the
critical value of $k_{0}$ decreases. For values of $k_{0}$ less than
$0.55k_{0}$ elastic scattering cannot occur, and the system should
be stable for all atom numbers. Dynamic instabilities of condensates
in moving lattices were recently observed in
Ref.\cite{Fallani04,Sarlo05}. In Ref. \cite{Sarlo05}, the chemical
potential was a factor of 3 higher than in our experiment, leading
to a dynamic instability for all values of $k_{0}$ above
0.55$k_{L}$.  Although discrete momentum states could not be
observed in those experiments, it is likely that the mechanism for
the dynamic instability is self-seeded parametric amplification.

In conclusion, we have demonstrated a matter wave analogue of both
optical parametric generation and optical parametric amplification
using a condensate moving in a one-dimensional optical lattice.
The optical lattice modified the dispersion curve and ensured
phase matching. If the separation of the phase matched momentum
states becomes less than the speed of sound, a condensate will
self-seed the process and become dynamically unstable.

The authors thank K. M{\o}lmer for pointing out this physical
process to us, M. Inguscio for insightful discussions, M. Saba for
insightful discussions and a critical reading of the manuscript, and
P. Medley for experimental assistance. This work was supported by
NSF and ARO.

\end{document}